\documentclass[a4paper,11pt]{article}
\usepackage{jinstpub} 
\usepackage{microtype}
\usepackage{subfig}
\usepackage{gensymb}

\title{Improving the light collection efficiency of silicon photomultipliers through the use of metalenses}

\author[a,1]{A.A.\ Loya Villalpando,\note{Now at the California Institute of Technology, Pasadena, CA, United States.}}
\author[a,2]{J.~Mart\'in-Albo,\note{Now at the Instituto de F\'isica Corpuscular (IFIC), Valencia, Spain.}}
\author[b]{W.T.~Chen,}
\author[a]{R.~Guenette,}
\author[a]{C.~Lego,}
\author[b,c]{J.S.~Park,}
\author[b]{F.~Capasso}

\affiliation[a]{Department of Physics, Harvard University, Cambridge, MA 02138, USA}
\affiliation[b]{John A.\ Paulson School of Engineering and Applied Sciences, Harvard University, Cambridge, MA 02138, USA}
\affiliation[c]{Nanophotonics Research Center, Korea Institute of Science and Technology (KIST), Seoul 02792, Republic of Korea}

\emailAdd{aloyavil@caltech.edu}

\abstract{Metalenses are optical devices that implement nanostructures as phase shifters to focus incident light. Their compactness and simple fabrication make them a potential cost-effective solution for increasing light collection efficiency in particle detectors with limited photosensitive area coverage. Here we report on the characterization and performance of metalenses in increasing the light collection efficiency of silicon photomultipliers (SiPM) of various sizes using an LED of 630~nm, and find a six to seven-fold increase in signal for a $1.3\times1.3~\mathrm{mm}^2$ SiPM when coupled with a 10-mm-diameter metalens manufactured using deep ultraviolet stepper lithography. Such improvements could be valuable for future generations of particle detectors, particularly those employed in rare-event searches such as dark matter and neutrinoless double beta decay.}


\arxivnumber{2007.06678} 

\begin{document}
\maketitle
\flushbottom

\section{Introduction}
Silicon photomultipliers (SiPMs) have become increasingly popular in the field of experimental particle physics due to their compactness, low fabrication cost, radio-purity, high gain at low bias voltage, high photon detection efficiency (PDE), fast timing, single photoelectron resolution, and insensitivity to magnetic fields. All of these qualities allow SiPMs to directly compete with or surpass the performance of conventional photomultiplier tubes (PMTs). A variety of rare-event experiments, such as those searching for neutrinoless double-beta decay \cite{Monrabal:2018xlr,Alvarez:2012sma,Albert:2017hjq} or direct detection of dark matter \cite{Aalseth:2020nwt,Aalbers:2016jon} employ SiPMs for tracking and calorimetry.

The design of any detector necessarily involves a trade-off between the desired performance, its simplicity of realisation and its ultimate cost. Using fewer and smaller SiPMs can significantly reduce the cost and complexity of a detector, but the decrease in photosensitive area coverage usually translates to degradation in tracking and calorimetry. One way to increase the light collection area while maintaining the advantages of using a low number of small SiPMs may be to couple each sensor to a lens of larger diameter. However, ordinary refractive lenses can be bulky and thus difficult to incorporate into the design of a particle physics detector. Metalenses \cite{Khorasaninejad1190}, a type of diffractive optical components that employ nanostructures as phase shifters to focus incident light \cite{Kildishev1232009,Yu333}, may be more suitable for this application. First, metalenses are compact and can be manufactured with radiopure materials, which is of high importance to many particle physics experiments. Furthermore, metalenses can be optimized for specific light wavelengths and have been shown to work with high efficiency above wavelengths of 260 nm \cite{1514255} with the possibility to extend to shorter wavelengths using fluorides. Lastly, they are fabricated with matured fabrication such as deep ultraviolet projection lithography, allowing mass production at low cost.  

Here we explore the possibility of coupling metalenses to SiPMs for their integration in particle detectors. In Section~\ref{sec:metalenses}, an introduction to metalenses, their working principle, and light transmission characterization results are provided. In Section~\ref{sec:exp}, the experimental design is introduced and results of light collection measurements by SiPMs with and without the use of metalenses are presented. These results are qualitatively validated by measurements of the transmitted light profile of our metalenses. In Section~\ref{sec:sec_conclusion}, conclusions are drawn and further studies are proposed.

\section{Metalenses} \label{sec:metalenses}

\subsection{Working principle}
\begin{figure}[tb]
\begin{minipage}{0.45\textwidth}
\centering
    \subfloat[Schematic representation of a metalens nanostructures. \label{fig:metalens_schematic}]{\includegraphics[height=0.67\textwidth]{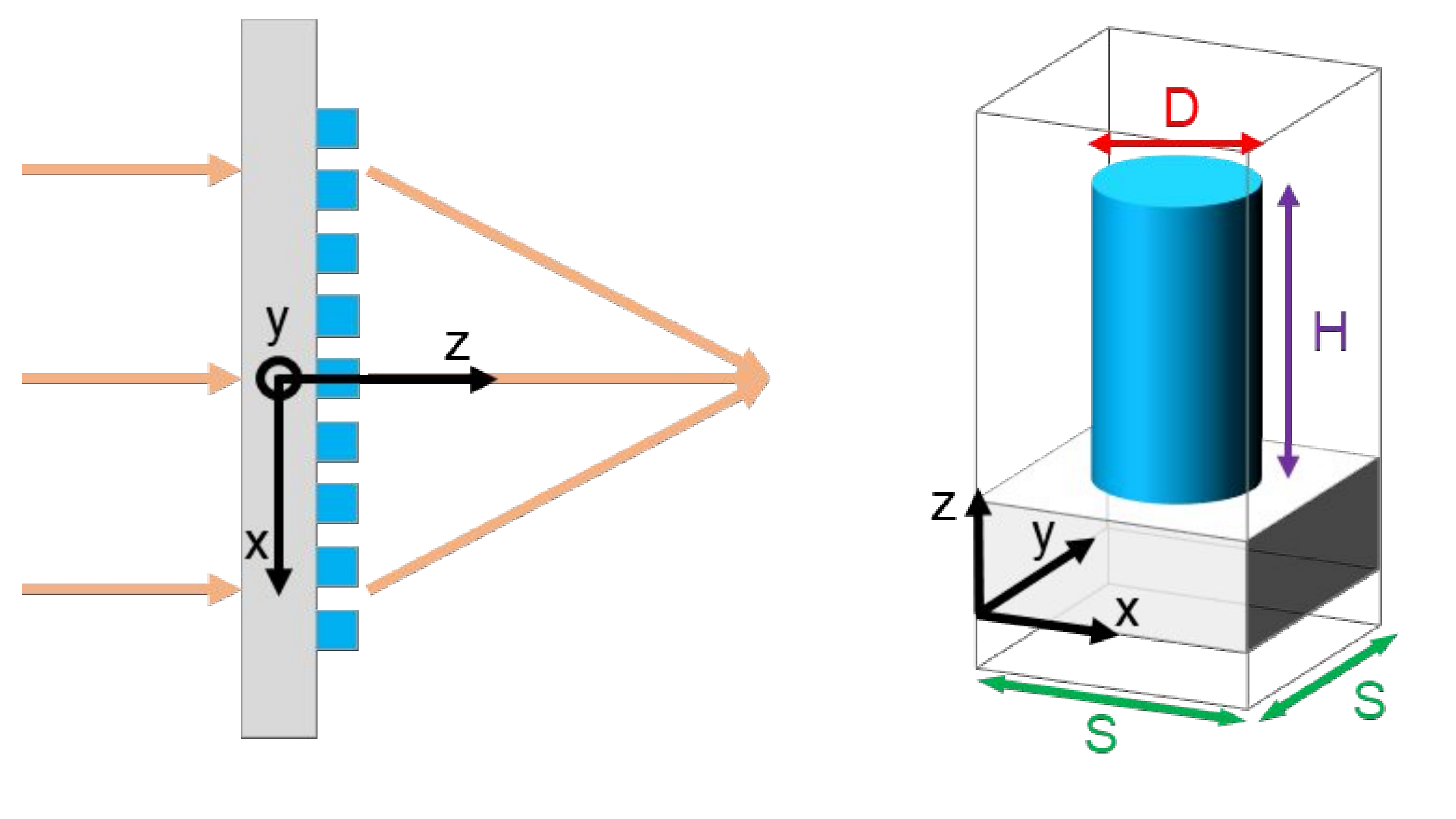}}
\end{minipage}
\begin{minipage}{0.5\textwidth}
\centering
\subfloat[SEM image of our metalens nanostructures.\label{fig:metalens_sem}]{\includegraphics[height=0.752\textwidth,angle=-90]{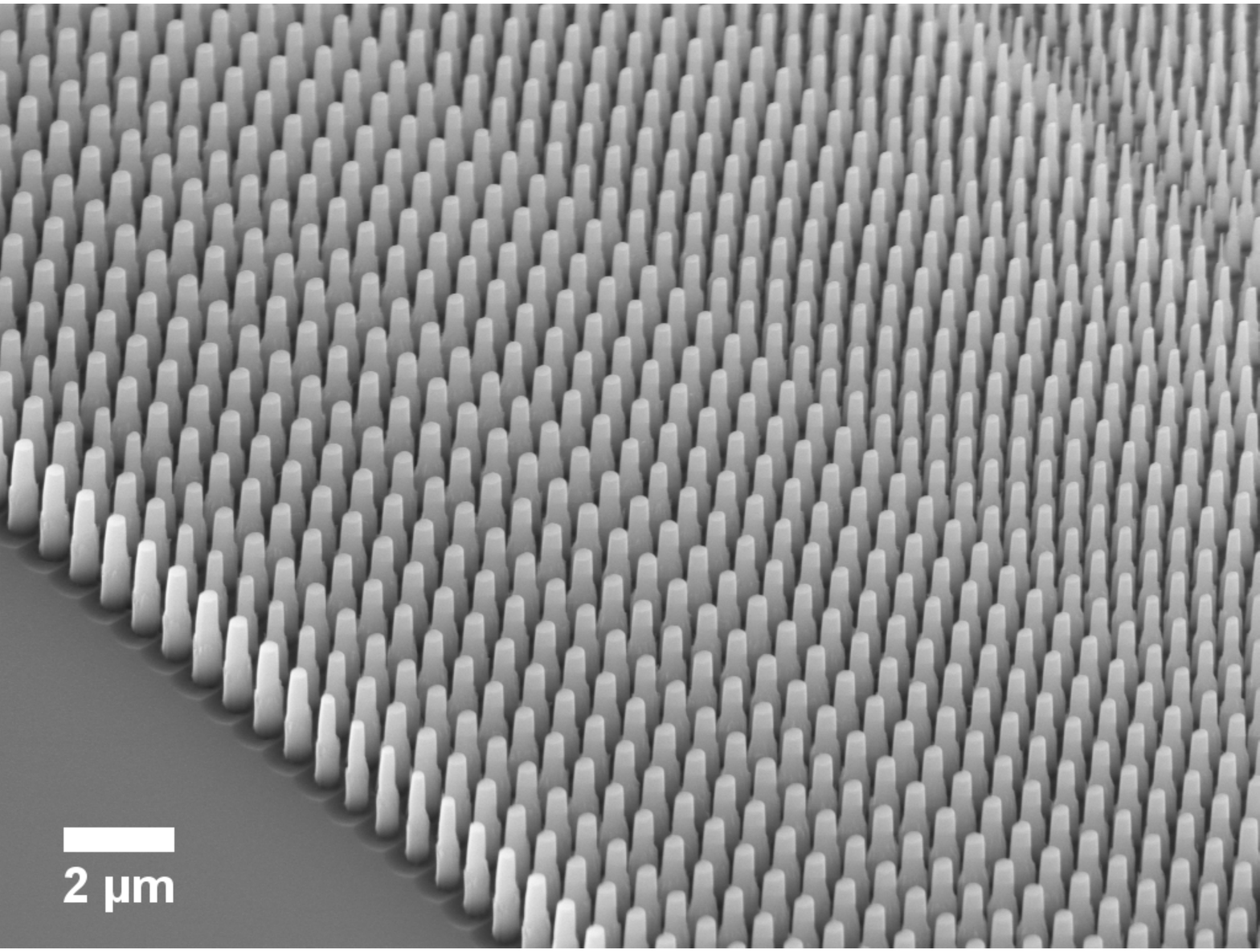}}
\end{minipage}
\begin{minipage}{\textwidth}
\centering
\subfloat[Array of identically fabricated metalenses composed of the nanostructures in (b). Each metalens has a diameter of $10~\mathrm{mm}$ and was designed at $630~\mathrm{nm}$ with a numerical aperture of $0.1$. \label{fig:metalens_array}]{\includegraphics[height=0.4\textwidth]{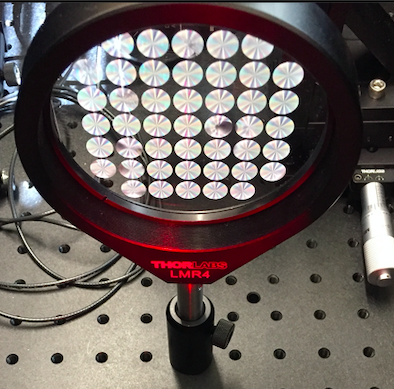}}
\end{minipage}
\caption{(a) Schematic of a metalens and its building blocks, SiO$_{2}$ nanopillars of varying diameters from $250~\mathrm{nm}$ to $600~\mathrm{nm}$. (b) A scanning electron microscope image of the edge of a metalens showing the nanopillars. These metalenses were fabricated by lithography followed by dry etching using trifluoromethane/argon (CHF$_{3}$/Ar) plasma. These nanopillars are 2-$\mu$m-height and tapered resulting from side wall etching. (c) Photograph of the metalens array used in this work.} 
\end{figure}

Metalenses consist of arrays of subwavelength-spaced dielectric nanostructures on a substrate serving as phase shifters that allow for unprecedented control over the wavefront of transmitted light. For instance, TiO$_{2}$, SiO$_{2}$ and GaN nanostructures of varying dimensions have been found to efficiently modulate phases in the visible range \cite{Devlin10473, Park19, doi:10.1021/acs.nanolett.7b03135}. Schematic and scanning electron microscope (SEM) images of a metalens and its nanostructures are shown in Figures \ref{fig:metalens_schematic} and \ref{fig:metalens_sem}, respectively. Due to their diffractive nature, metalenses produce multiple foci, each corresponding to a different diffraction order (see Figure~\ref{fig:foci_schematic}).

The circular pillars shown in Figure~\ref{fig:metalens_sem} were designed using a commercial finite difference time domain (FDTD) solver (\textsc{Lumerical Inc., Vancouver}). They possess different diameters resulting in different phase delays to mimic a focusing lens by implementing a target phase profile according to \cite{doi:10.1021/nl302516v}
\begin{equation}\label{eq:phase}
\phi_{nf}(r) = \frac{2\pi}{\lambda_{d}} \left(f - \sqrt{r^{2} + f^{2}} \right ),
\end{equation}
where $\lambda_{d}$ is the design wavelength, $r$ is the radial coordinate, and $f$ is the focal length. However, due to the resolution limitation of the deep ultraviolet stepper we used (\textsc{PAS5500/300C DUV} Stepper, \textsc{ASML}), these nanopillars cannot fully cover a range of $2\pi$ phase delay leading to multiple foci \cite{Park19}. The reason for having such multiple foci can be understood by treating the metalens as stitched gratings with spatially varying periodicities \cite{o2004diffractive}. When the circular pillars within a local periodicity cannot impart a linear phase change to incident light, the transmitted light diffracts to multiple grating orders resulting in the multiple foci. Additionally, the diffraction angle of each order follows the grating equation
\begin{equation}\label{eq:all}
n_{t}~\sin\theta_{t} - n_{i}~\sin\theta_{i} = \frac{n}{p} = \frac{\lambda_{o}}{2\pi}\frac{d\phi}{dr}\,,
\end{equation}
where $\theta_{i}$, $\theta_{t}$, $n_{i}$ and $n_{t}$ are the angle of incidence, angle of refraction, and the refractive indices of the incidence and transmission media, respectively; $\lambda_{0}$ is the vacuum wavelength of incident light; $n$, the corresponding diffraction order and $p$, the local periodicity of nanofins at location $r$ from the center of the metalens. Equations \eqref{eq:phase} and \eqref{eq:all} fully determine the propagation of transmitted light for each position of a metalens. The transmitted power that diffracts to each order can then be measured by a detector.

\subsection{Light transmission efficiency}

\begin{figure}[tb]
\centering
\includegraphics[scale=0.6]{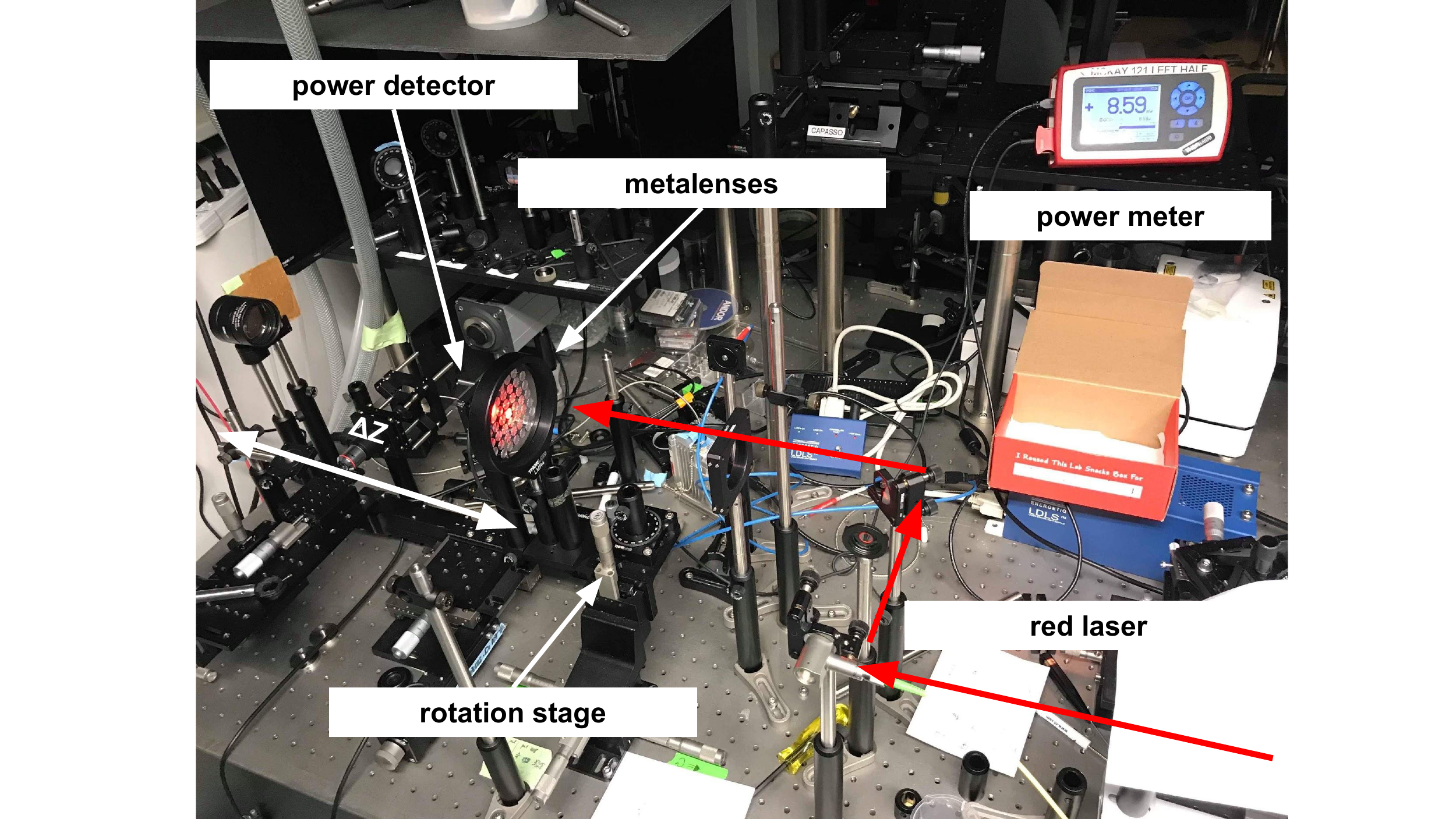}
\caption{ The setup used to perform light transmission efficiency measurements of metalenses. A red laser of variable diameter shines on a single metalens and the transmitted power is measured using a power detector at various distances from the metalens ($\Delta$Z). The laser, metalens and power detector were aligned along an optical axis in normal incidence. For another angular efficiency measurement, the metalens was mounted on the rotation stage.}
\label{fig:eff_setup}
\end{figure}

Figure~\ref{fig:metalens_array} shows an array of 45 identically designed all-glass circular metalenses fabricated on a single glass substrate, each with a design wavelength ($\lambda_{d}$) of 632.8~nm and a diameter of 10~mm \cite{Park19}. To better understand the functionality of these metalenses for their application in a particle detector, a few transmission efficiency measurements were carried out with the setup shown in Figure~\ref{fig:eff_setup}. From this point forward any mention of transmission efficiency refers to the fraction of transmitted power across the metalens divided by the total incident power on the metalens. For all transmission efficiency measurements, an unpolarized laser beam (\textsc{NKT SuperK EXTREME} paired with \textsc{SuperK VARIA}) with a $40~\mathrm{nm}$ bandwidth centered at $630~\mathrm{nm}$ was used. 

\begin{figure}[tb]
\centering
\subfloat[Metalens transmission efficiency vs distance from \newline the power sensor.\label{fig:lin_eff_results}]{\includegraphics[width=.50\textwidth]{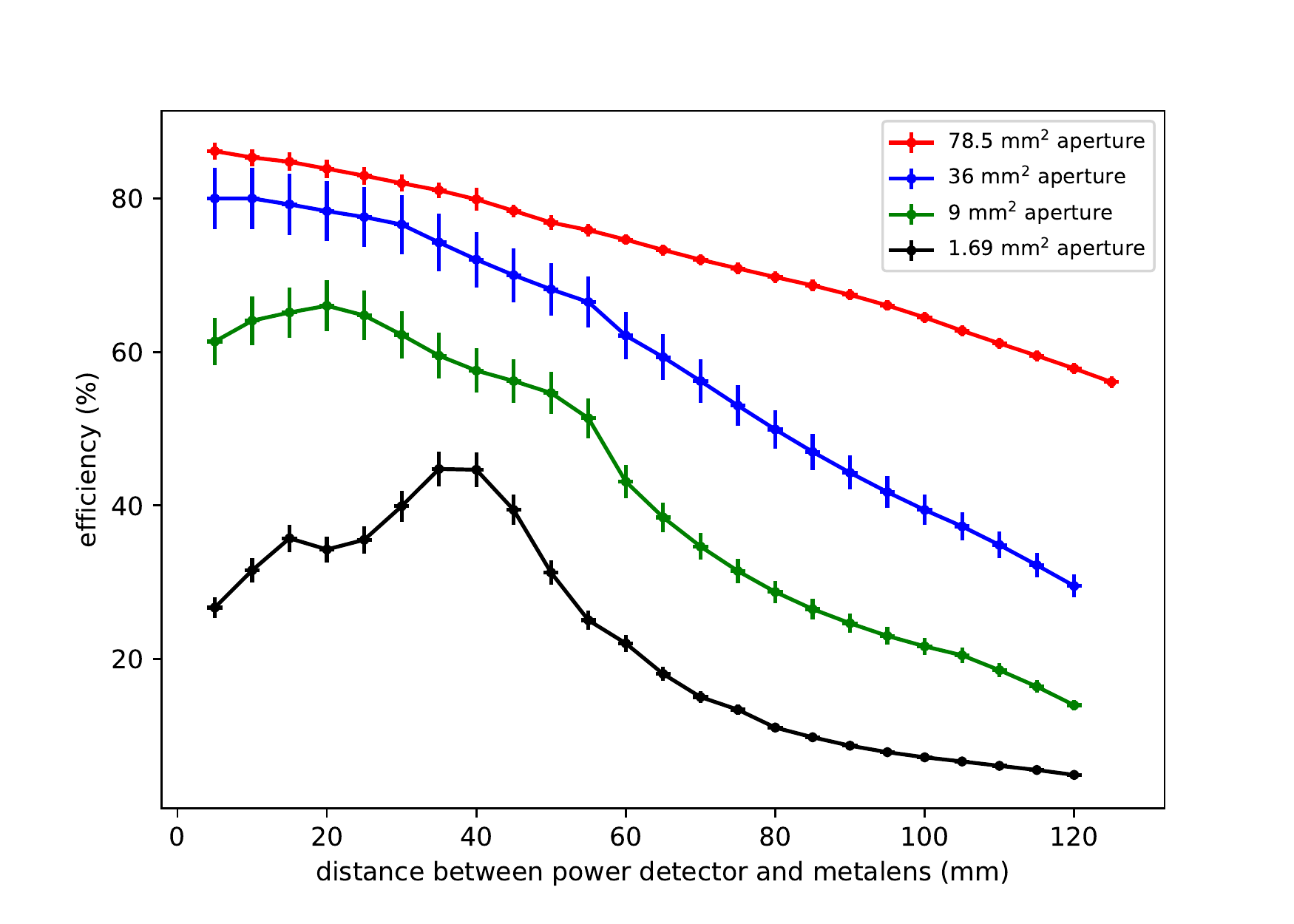}}
\subfloat[Metalens transmission efficiency vs incident beam angle.\label{fig:angular_results}]{\includegraphics[width=.50\textwidth]{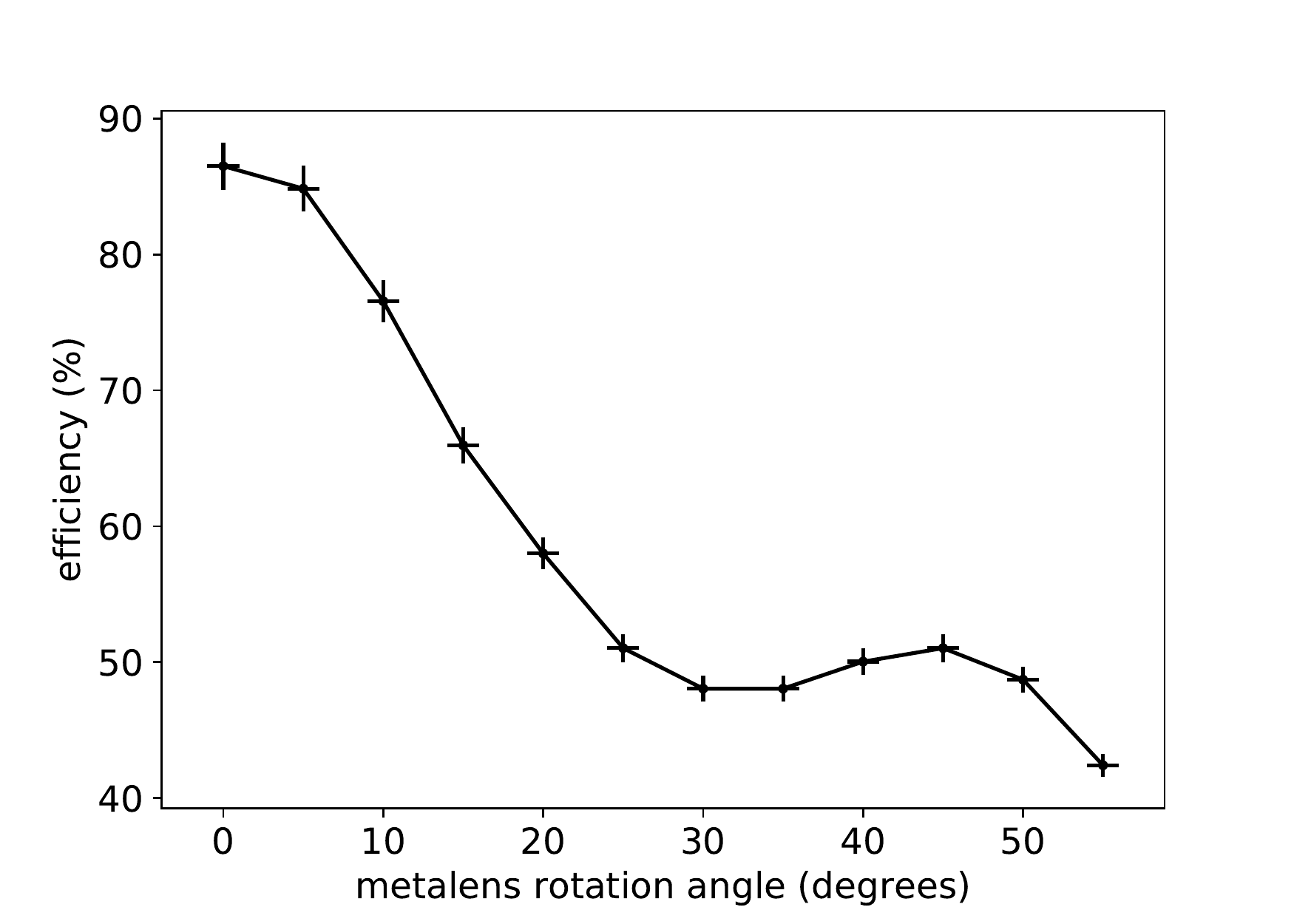}} \\
\subfloat[2D schematic representation of foci contributions to light transmitted by a metalens. \label{fig:foci_schematic}]{\includegraphics[width=.75\textwidth]{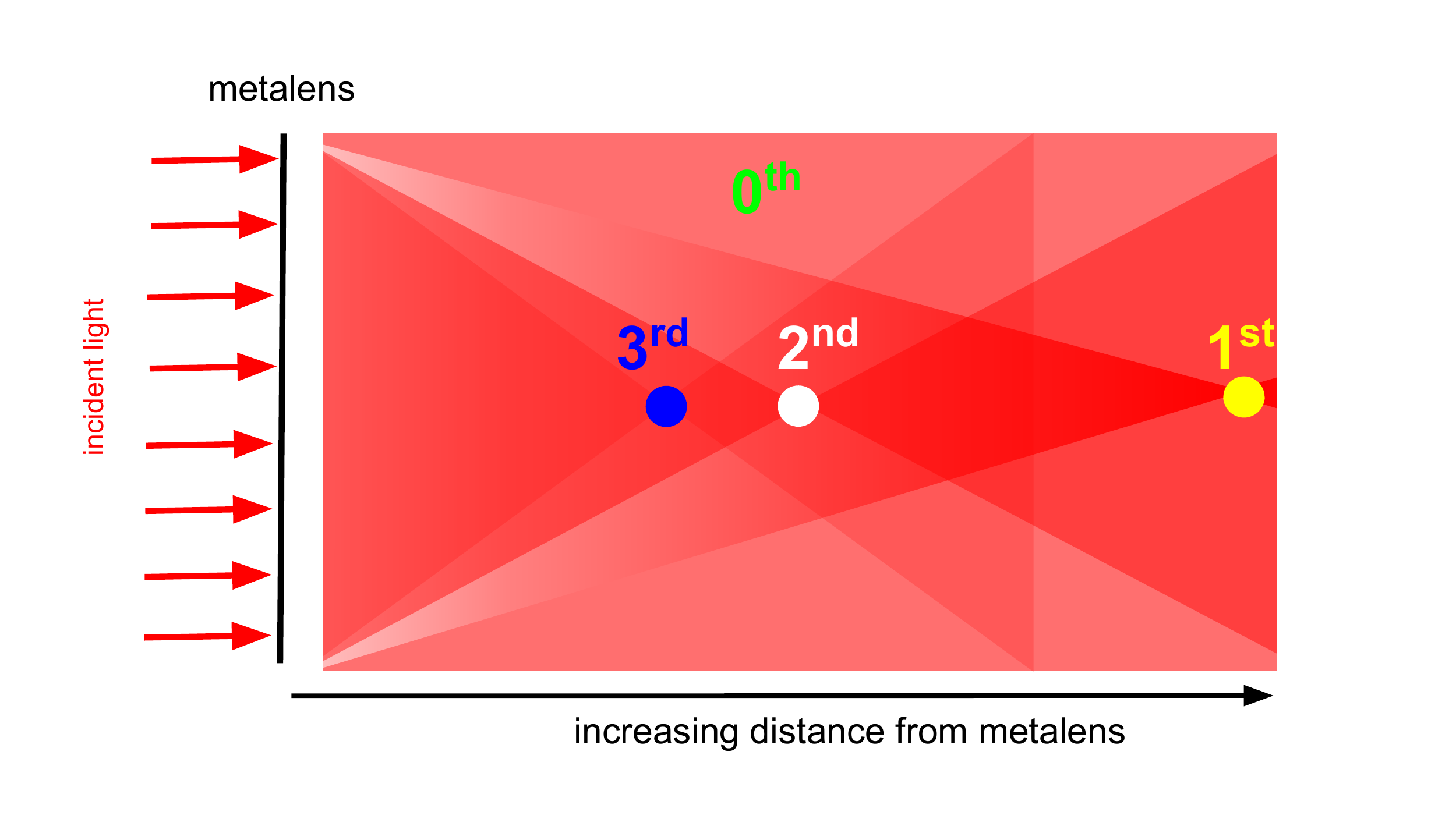}}
\caption{(a) Light transmission efficiency of metalenses using a power detector with various aperture sizes as a function of distance from the metalens. (b) Light transmission efficiency of metalenses as a function of the beam's incident angle. (c) Two-dimensional schematic representation of the different diffraction order contributions to the overall transmitted light by a metalens.}
\label{fig:characterization}
\end{figure}

First, the efficiency per diffraction order of the metalenses was measured to determine the relative contributions of each diffraction order to the foci (see Figure~\ref{fig:foci_schematic}). A collimated $\sim$2 mm beam passed through a $\sim$50 mm  focusing lens (\textsc{Thorlabs AC508-150-A-ML}, not shown in Figure~\ref{fig:eff_setup}) and was focused on a metalens, effectively yielding an incident plane wave. The beam scanned different positions $r$ from the center to the edge of the metalens in steps of 0.5 mm, and the amount of power transmitted to each diffraction order was measured using a power sensor (\textsc{Thorlabs S121C}) with a 10 mm circular aperture, which fully captured the diffracted beams. The procedure was repeated for three different metalenses and average efficiencies of $(20\pm2)\%$,  $(36\pm1)\%$, $(11\pm2)\%$, and $(4\pm1)\%$ were found for the 0$^{\text{th}}$, 1$^{\text{st}}$, 2$^{\text{nd}}$ and 3$^{\text{rd}}$ diffraction orders averaged over all values of $r$, respectively. The center of the metalens ($r < 0.5$~mm) does not provide a phase gradient and allows light to pass through without being diffracted, reaching an average efficiency of $(80.3\pm0.6)\%$. The fact that the 1$^{\text{st}}$ order focus was found to have the highest efficiency is compatible with design specifications.

Second, we measured how power varies along the optical axis of the metalens using the setup shown in Figure~\ref{fig:eff_setup}. This measurement is important for understanding the amount of light transmitted through our metalenses to a specific SiPM area placed behind the metalens. As seen in Figure~\ref{fig:eff_setup}, the center metalens was illuminated by a collimated laser beam. The power detector, mounted together with a tunable circular aperture, was placed behind the metalens and was precisely moved by a motorized stage along the z-direction. We recorded power values using aperture areas of 78.5 mm$^{2}$, 36 mm$^{2}$, 9 mm$^{2}$, 1.69 mm$^{2}$ (matching the area of the metalens and the active area of the three SiPMs used in Section~\ref{sec:experiment_design}) and normalized them to the power of the incident laser beam. The measured efficiencies are shown in Figure~\ref{fig:lin_eff_results}, which can be interpreted using the schematic shown in Figure~\ref{fig:foci_schematic}. In general, moving away from the metalens results in a decrease in power detection due to the divergence of the foci beyond the power sensor aperture, as seen when using a power sensor aperture of 78.5 mm$^{2}$ (red curve in Figure \ref{fig:lin_eff_results}). The rate of decrease is limited by the 0$^{\text{th}}$ order contribution, which is the second most dominant. The contributions of the different diffraction orders are most evident when using an aperture of 1.69 mm$^{2}$  (black curve in Figure \ref{fig:lin_eff_results}). An increase in efficiency is observed as the power sensor moves towards the 1$^{\text{st}}$ order focus with increasing contributions from all three foci, and begins to decrease at a distance of $\sim$40 mm due to a reduction in the contribution from the the 3$^{\text{rd}}$ and 2$^{\text{nd}}$ foci. This measurement indicates that there is an optimal location to place a SiPM behind a metalens for maximum light collection.

Lastly, we measured the light transmission efficiency of the 1$^{\text{st}}$ order focus (diffraction order with highest transmitted power) as a function of the beam's angle of incidence, which is an important measure for the application of metalenses in a particle detector, as light would arrive at the interface of a metalens from a wide range of angles. A collimated beam of $\sim$5 mm was pointed at the center of a single metalens mounted on a manual rotation stage (\textsc{Thorlabs RP01}) while the metalens was rotated in steps of 0.5\degree. The collimated beam diameter was set to be less than the full width of a metalens in order to capture all of the transmitted beam, which becomes increasingly elliptical with increasing incident angle. The transmitted power as a function of rotation angle of the metalens was measured using a 10 mm power sensor aperture and the results are shown in Figure~\ref{fig:angular_results}. We find an average angular transmission efficiency of $(59\pm1)\%$ between 0\degree and 55\degree. It is worth mentioning that metalenses have better angular efficiency compared to conventional saw-tooth Fresnel lenses because the latter suffer from shadow effects \cite{doi:10.1021/acsphotonics.9b00221}.

\section{Light collection of SiPMs with metalenses} \label{sec:exp}

\subsection{Experimental design}\label{sec:experiment_design}
\begin{figure}
\includegraphics[width=.5\textwidth]{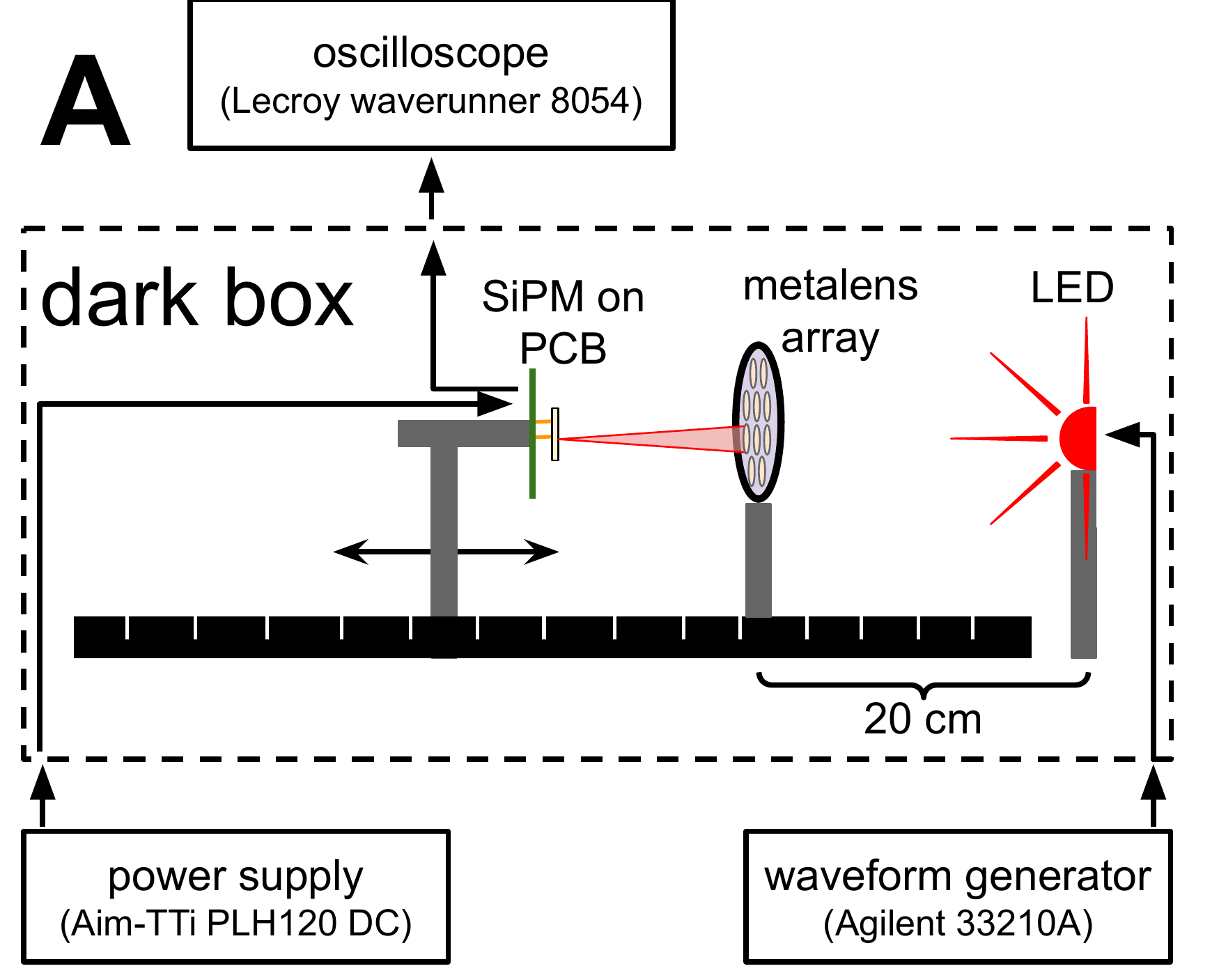}
\includegraphics[width=.5\textwidth]{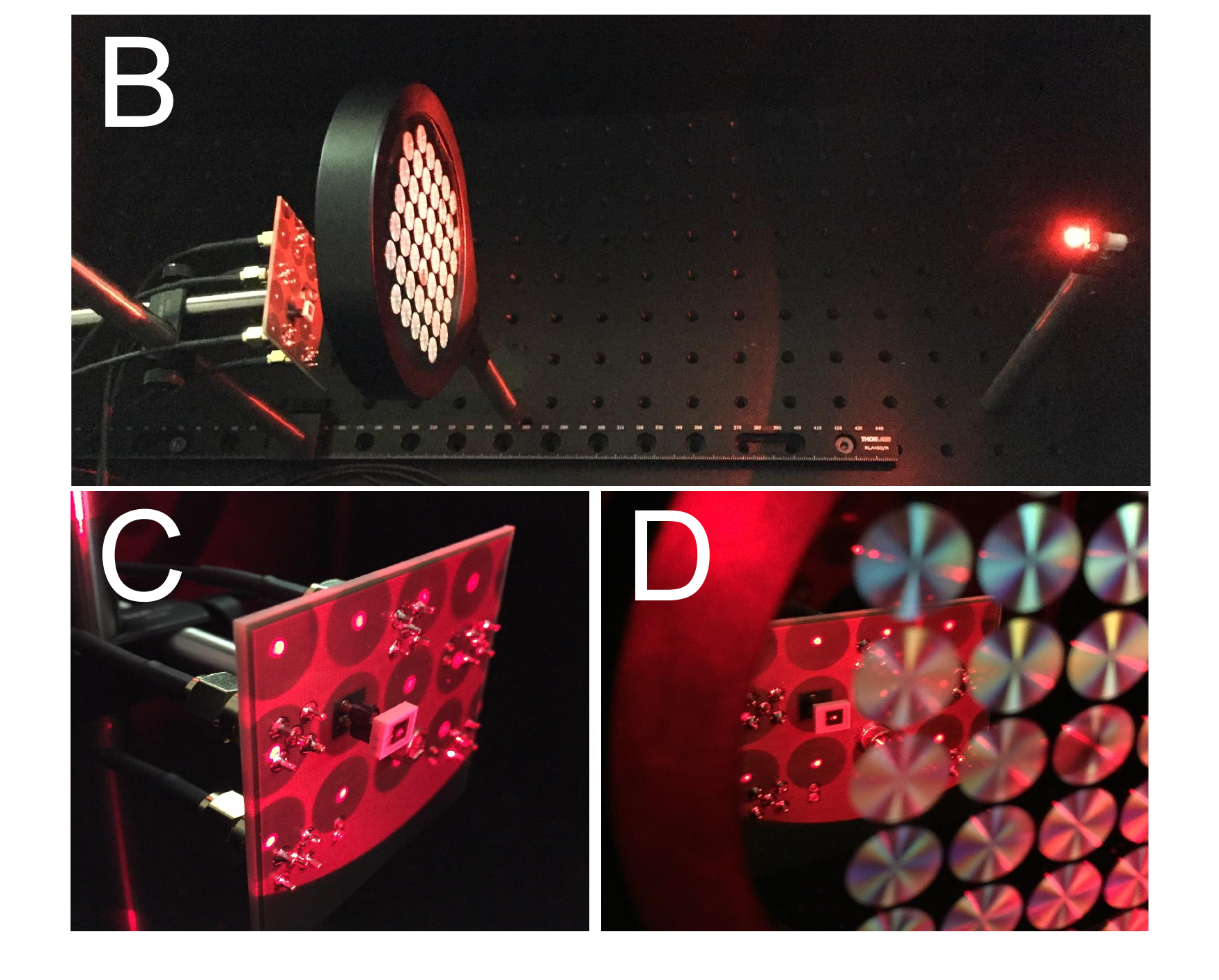}
\caption{(A) Schematic and (B) photograph of the setup used to quantify the light collection of SiPMs with and without the use of a metalens. A red LED was placed $20~\mathrm{cm}$ away from the metalens array and a SiPM was aligned with the focal axis of a single metalens. The SiPM signal was measured as a function of distance from the metalens array location. (c) and (D) show the spot size created on the $3\times3~\mathrm{mm}^{2}$ SiPM when it is located slightly away from and at the location of the 1$^{st}$ focus, respectively.}
\label{fig:black_box_setup}
\end{figure}

A schematic and photographs of the setup used to measure the SiPM signals with and without the use of metalenses are shown in Figure~\ref{fig:black_box_setup}. It consists of a black optical enclosure with a honeycomb breadboard. A light emitting diode (LED) with a wavelength peaked around 630 nm (\textsc{Thorlabs LED630L}) is mounted at one end of the breadboard using an optical pillar post and an LED mount (\textsc{Thorlabs LEDMF}). Identical LED settings were used for all SiPMs when coupled and not coupled with a metalens. The LED was driven by pulses of 1.8 V, 40 ns width, and 5 ns rise time at a frequency of 500 Hz using a waveform generator (\textsc{Agilent 33210A}). The pulse width was chosen to be shorter than the average recovery time of each SiPMs' microcells (60 to 100 ns), while the frequency was low enough to allow full recovery of all microcells between consecutive pulses. The amplitude was chosen to generate enough photons for signal detection without saturating any of the SiPMs. The array of metalenses was mounted $20~\mathrm{cm}$ away from the LED using an optical pillar and a fixed lens mount (\textsc{Thorlabs LMR10}) (see Figure \ref{fig:metalens_array}). On the other side of the metalens array a custom printed circuit board (PCB) holding a single SiPM was mounted onto a dovetail optical rail (\textsc{Thorlabs RLA300/M}) using a couple of optical posts and a right angle clamp (\textsc{Thorlabs RA90}). The optical rail sat close and parallel to the central axis of the LED-metalens-SiPM system and served to identify the relative location of the metalenses and the SiPMs. The SiPMs were biased to 54.6 V using a DC power supply (\textsc{Aim-TTi PLH120}) and their signals were recorded using an oscilloscope (\textsc{Lecroy waverunner 8054}), which calculated the area of the SiPM pulses and the standard deviation of the distribution.

The amount of light collected by individual SiPMs of various sizes is compared when coupled to a single metalens with the amount of light collected by the SiPMs without using a metalens while keeping all other variables constant. SiPMs from \textsc{Hamamatsu (S13360)} with photosensitive areas of $1.3\times1.3~\mathrm{mm}^{2}$, $3\times3~\mathrm{mm}^{2}$, and $6\times6~\mathrm{mm}^{2}$ (667, 3600 and 14400 pixels, respectively) were used. All three sensors have a pixel pitch of 50 $\mu$m and a photo detection efficiency (PDE) of $\sim$21\% at 632 nm. Each SiPM's active area was aligned with the focal axis of a single metalens and the signals generated by the SiPM at different distances from the metalens location were recorded. The SiPMs were then illuminated while located at the same locations,but with the metalens removed to serve as a reference measurement.

\subsection{Focused light profile}
\begin{figure}[tb]
\centering
\subfloat[Spot size profiling setup.\label{fig:spot_measure}]{\includegraphics[width=.425\textwidth]{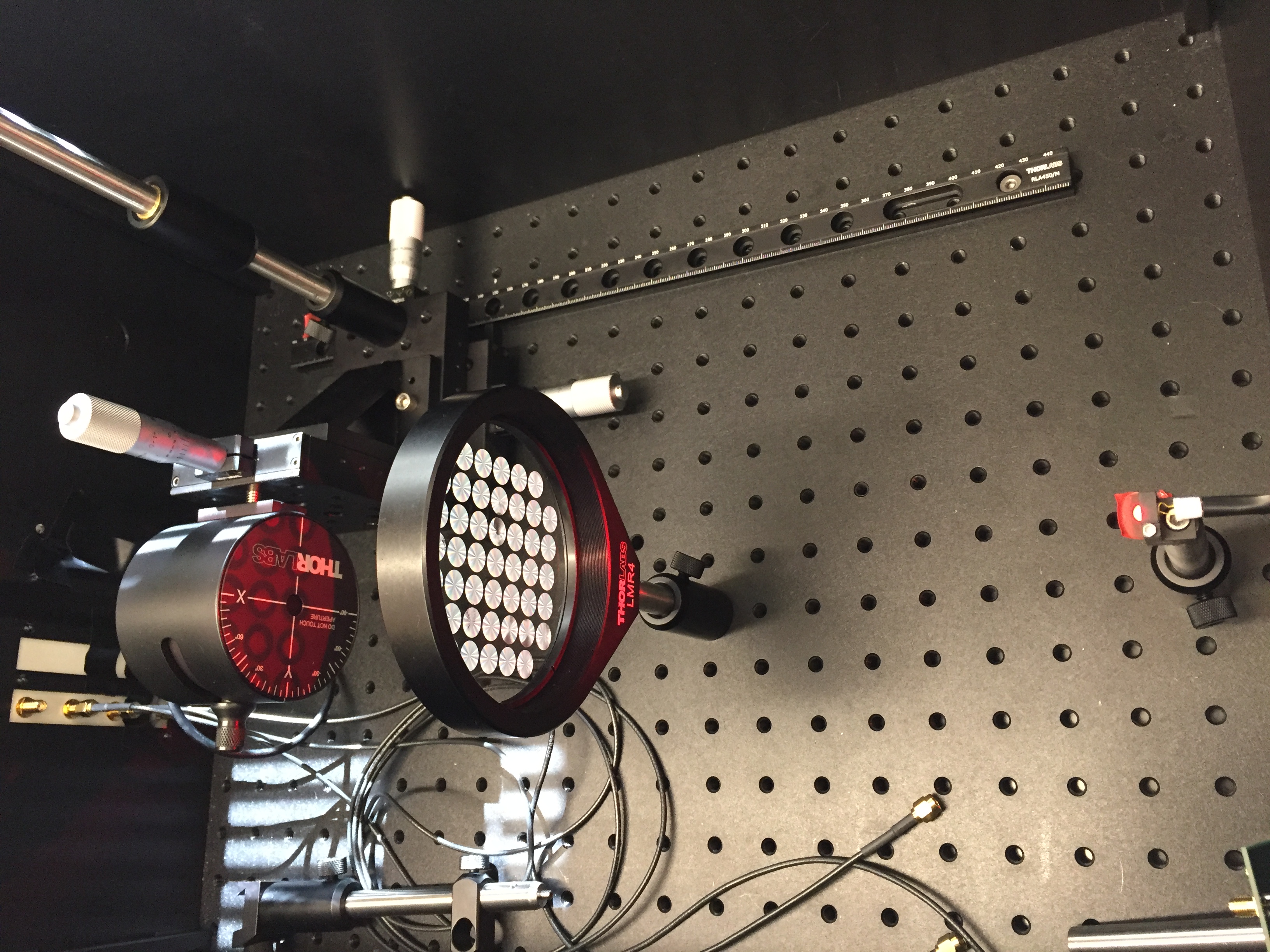}} \quad
\subfloat[Spot size as a function of distance from the metalens. \label{fig:spot_results}]{\includegraphics[width=.525\textwidth]{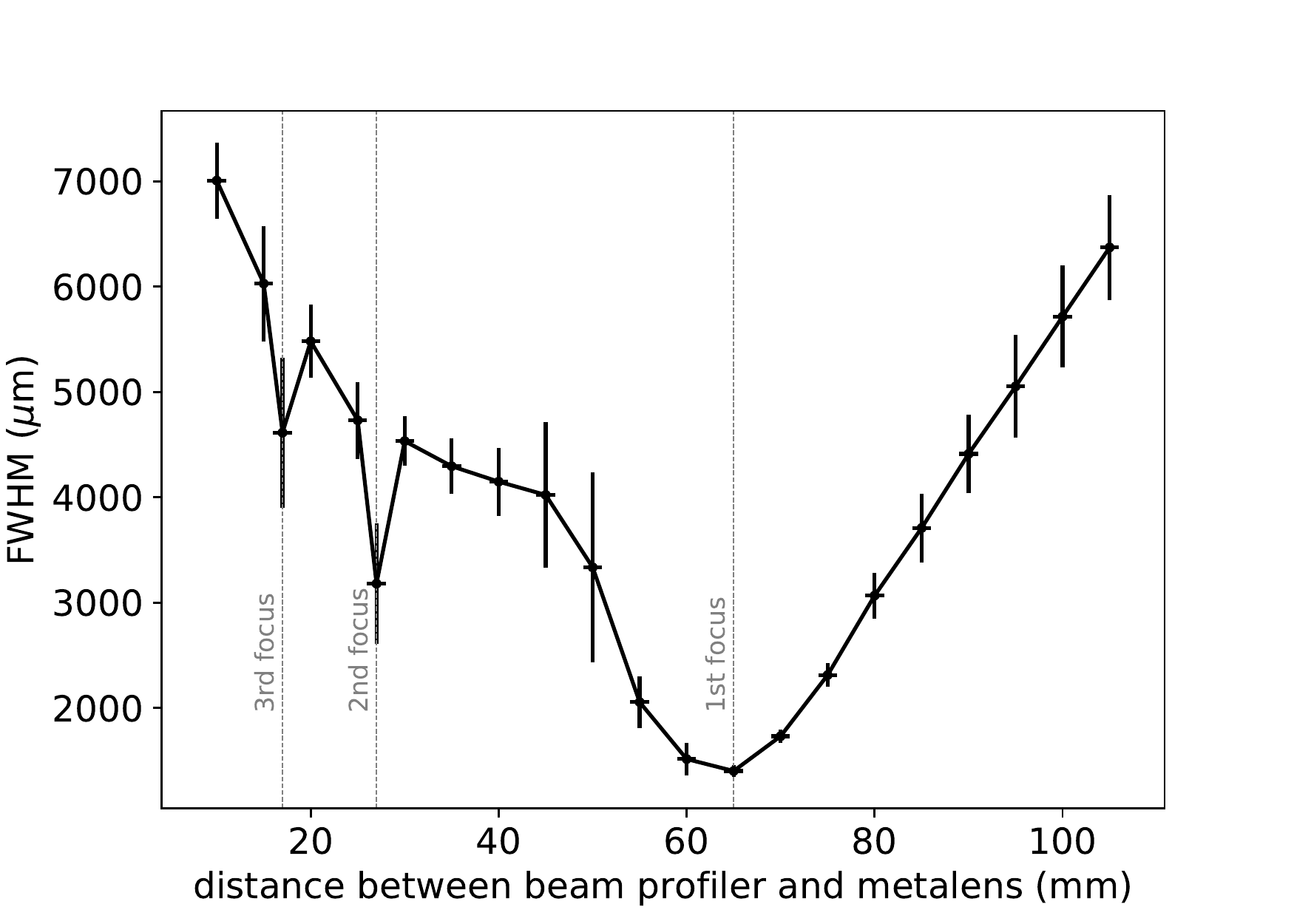}}
\caption{(a) Setup and (b) results for the measurements of the spot size produced by the metalens. A beam profiler was aligned with the optical axis of a single metalens that was illuminated by a red LED as in Figure~\ref{fig:black_box_setup}. The beam profiler was placed at multiple distances from the metalens and the FWHM of the beam was recorded. The results in (b) show the 1$^\text{{st}}$, 2$^{\text{nd}}$, and 3$^{\text{rd}}$ measured focus at a distance around $65~\mathrm{mm}$, $27~\mathrm{mm}$, and $17~\mathrm{mm}$ away from the metalens, respectively. The spot size at the location of the 1$^\text{{st}}$ focus was found to be 1.40$\pm$0.06~mm. }
\label{fig:spot}
\end{figure}

\begin{figure}[tb]
\centering
\subfloat[2D transmitted light profile. \label{fig:2d_intensity}]{\includegraphics[height=7cm]{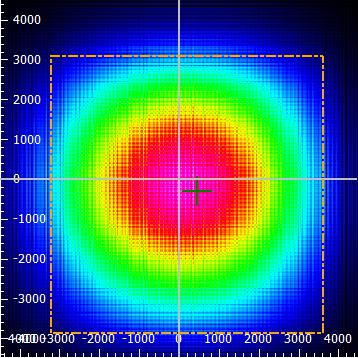}} \quad
\subfloat[3D transmitted light profile.\label{fig:3d_intensity}]{\includegraphics[height=7cm]{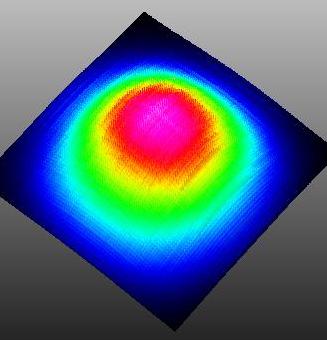}}
\\
\subfloat[X-profile intensity distribution. \label{fig:2d_x}]{\includegraphics[width=0.475\textwidth]{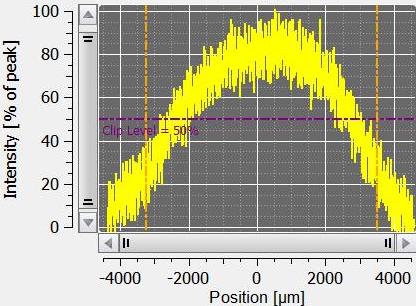}} \quad
\subfloat[Y-profile intensity distribution.
\label{fig:2d_y}]{\includegraphics[width=0.475\textwidth]{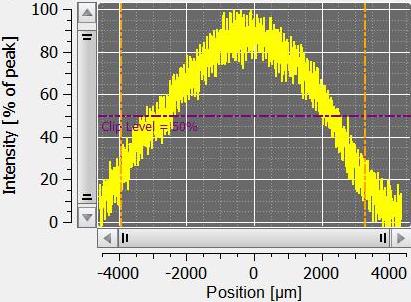}}
\caption{Images from the beam profiler software with the profiler located 20 mm away from the metalens. (a) 2D profile of the light transmitted by a metalens with the spot size enclosed by the orange dash-dotted rectangle. The color gradient represents areas of higher intensity in pink/red near the center and of lower intensity in green/blue further from the center. Axes units are shown in $\mu$m. (b) 3D profile. (c) Single snapshot of the x-profile intensity distribution. (d) Single snapshot of the y-profile intensity distribution. The spot size boundaries are marked by the vertical orange dash-dotted lines, while the FWHM is marked by the horizontal purple dash-dotted lines.}
\label{fig:beam_profiles}
\end{figure}

SiPMs are pixelated sensors which produce signals proportional to the number of pixels that fire, which depends on their PDE and therefore on the distribution of light on the pixels. We define the spot size as the area enclosing light intensity greater than 13.5\% ($1/e^{2}$) of the peak intensity. To understand the SiPM signal dependence on spot size when using a metalens, the spot size diameter was measured using a beam profiler (\textsc{Thorlabs BP209-VIS}) in the same setup described in Section~\ref{sec:experiment_design} by replacing the SiPM with the profiler, as shown in Figure \ref{fig:spot_measure}. The profiler measures the distribution of the light intensity scaled to the maximum intensity detected in two directions (x,y) independently. Images from the beam profiler software are shown in Figure \ref{fig:beam_profiles}. Because we found the spot size to be over 98\% symmetric in x and y, we report the FWHM of the average spot size diameter at every distance from the metalens, as shown in Figure \ref{fig:spot_results}. At each distance, 100 samples of the profile in each dimension (x and y) were recorded and averaged. This was repeated for 3 different metalenses and the errors reported on the FWHM are dominated by variations in the alignment of the components, while a 2 mm uncertainty is estimated on the distance between the profiler and the metalenses. Three local minima are observed in Figure \ref{fig:spot_results} at 17 mm, 27 mm and 65 mm, corresponding to the measured 3$^{\text{rd}}$, 2$^{\text{nd}}$ and 1$^{\text{st}}$ focal points with increasing distance from the metalens, respectively.

\subsection{Results}
\begin{figure}[tb]
\centering
\subfloat[SiPMs' signals when coupled to a metalens.\label{fig:sipms_with_lens}]{\includegraphics[width=0.49\textwidth]{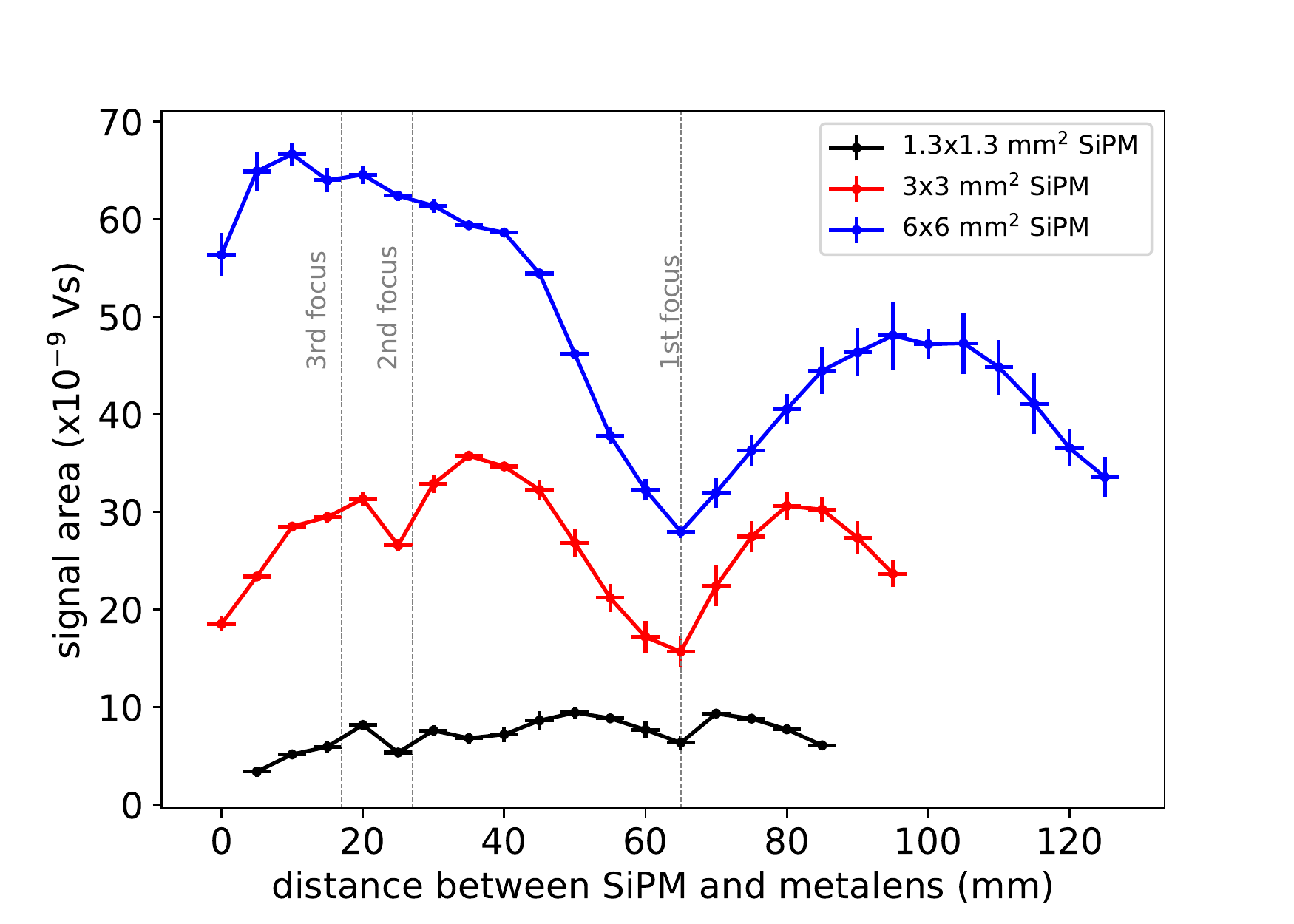}}
\subfloat[SiPMs' signals without using a metalens.\label{fig:sipms_without_lens}]{\includegraphics[width=0.49\textwidth]{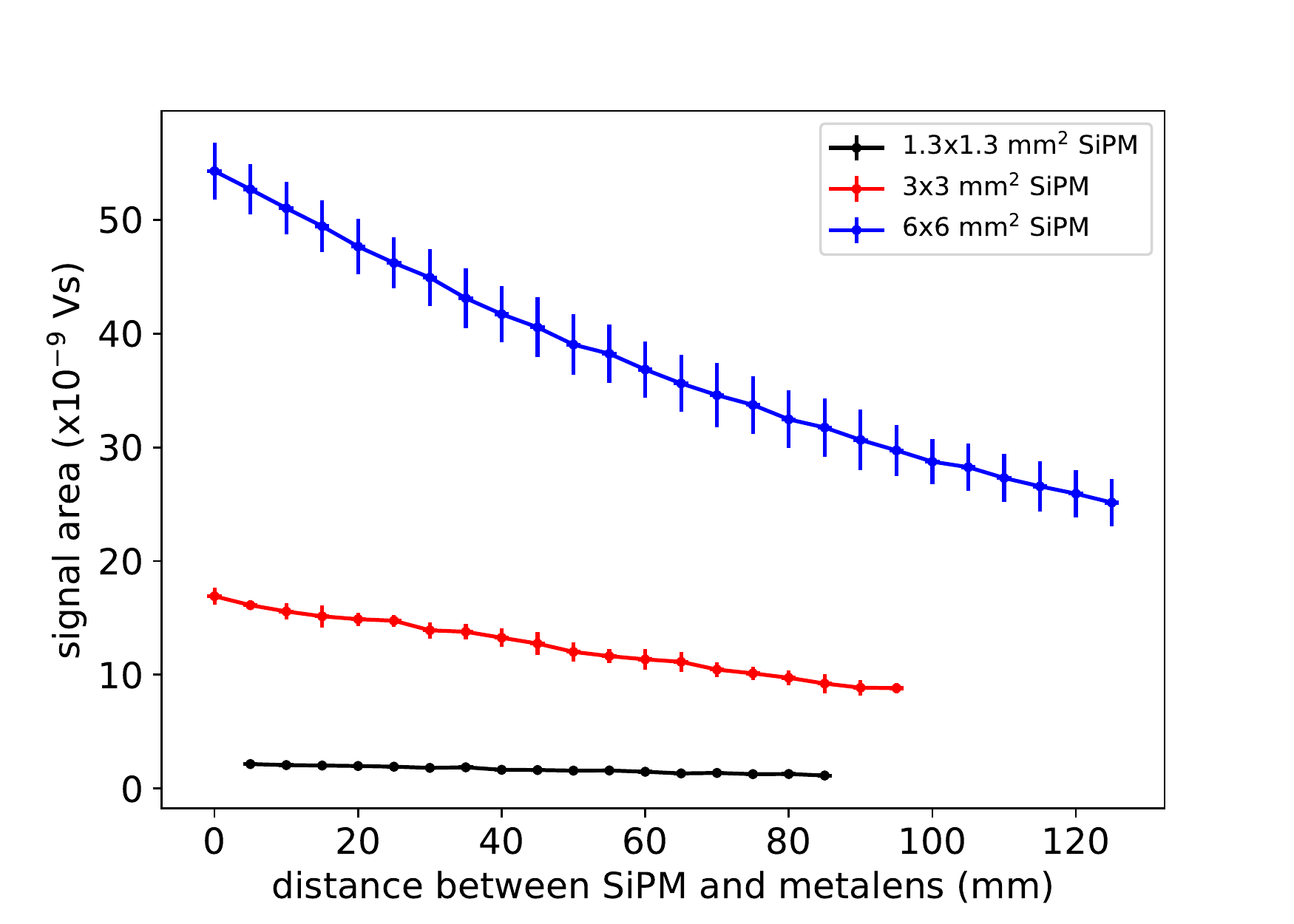}} \\
\subfloat[SiPMs' signal multiplication factor.  \label{fig:sipms_gain}]{\includegraphics[width=0.49\textwidth]{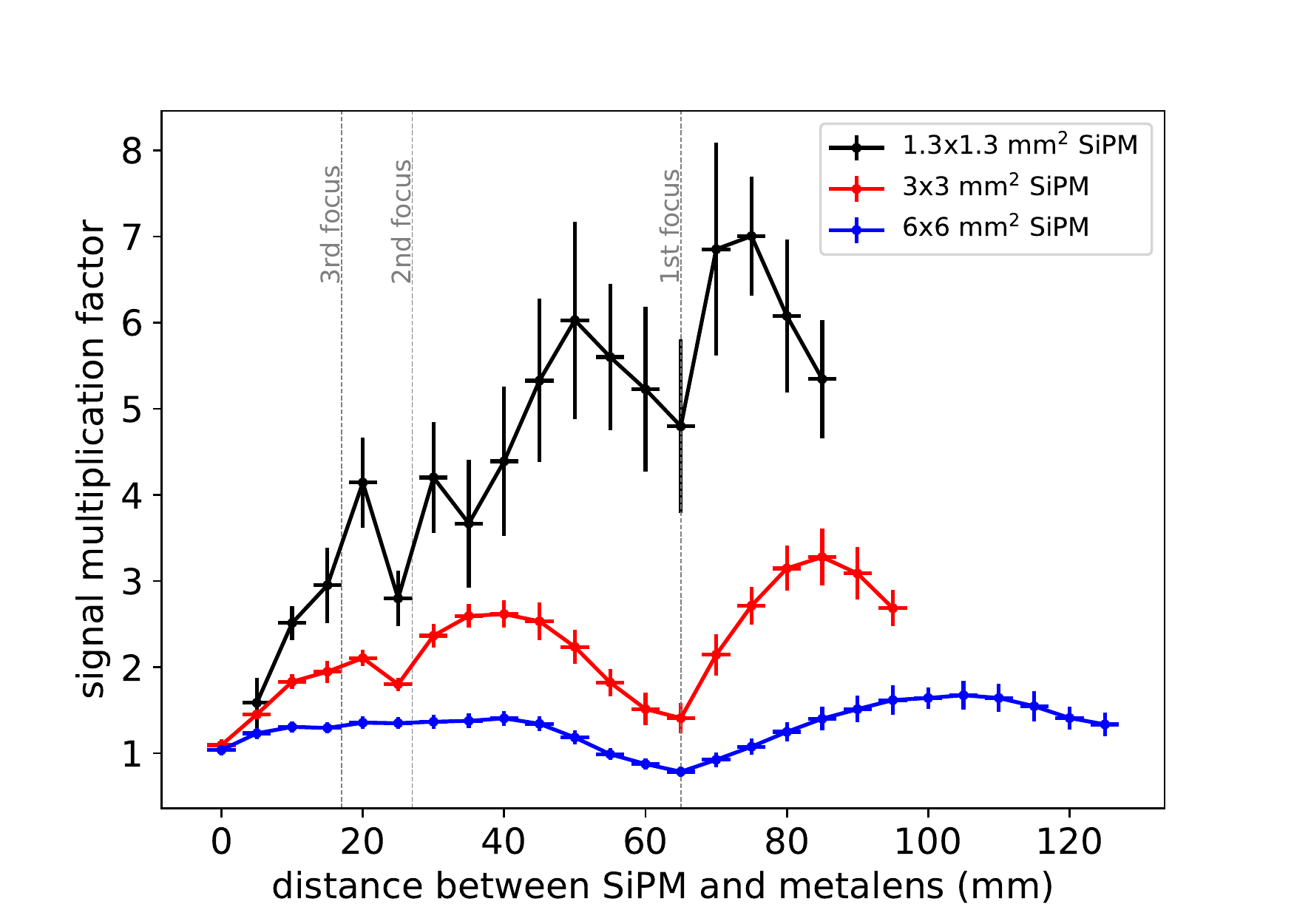}}
\caption{SiPM signals collected as a function of distance from the metalens (a) using a metalens and (b) without using a metalens. (c) The signal multiplication factor as a function of distance from the metalens. Descriptions of the signal structures are provided in the text.}
\label{fig:sipm_signals}
\end{figure}

Figures \ref{fig:sipms_with_lens} and \ref{fig:sipms_without_lens} show the amount of signal recorded by the different size SiPMs as a function of distance from the metalens with and without using the metalens, respectively. In both scenarios, it is observed that the larger the SiPM's active area, the larger the signal produced at any distance from the metalens, a result only dependent on the number of pixels of each SiPM. The signals recorded without using a metalens show the expected inverse squared distance dependence for all three sensors. 

As seen in Figure \ref{fig:sipms_with_lens}, when using the metalens, an asymmetric two-arch profile is observed in the signal recorded, which is qualitatively understood by the size of each SiPM's active area and the transmitted light profile (spot size diameter and intensity distribution). In Figure \ref{fig:sipms_with_lens}, all three curves show a dip around 65 mm (D1), which corresponds to the distance associated with the minimal spot size in Figure \ref{fig:spot_results}. At this distance, most of the spot's intensity is fully contained within each SiPM's active area. Under this condition, the light is focused on a number of pixels less than the total number of pixels in each SiPM and therefore a relatively low signal is produced. While the focusing is the same on all three SiPMs at this distance, the amount of signal increases with increasing SiPM size. This is due to an increase in the number of pixels on the SiPM since the lower intensity  portions of the spot distribution also induce signals on pixels far from the center of each SiPM, albeit with a lower probability that is dependent on the SiPMs' PDE. Moving away from D1 in either direction results in an increase in signal for all three SiPMs, which is explained by the increase in spot size that results in an increased coverage of each SiPM's active area. As the spot size increases, the spot intensity is spread out over a wider diameter, increasing the number of pixels away from the center of the SiPMs' active area that contribute to the signal. The maximum value of each arch is reached when the spot is focused just enough to fully cover a SiPM's active area. This can be deduced from Figures \ref{fig:sipms_with_lens} and \ref{fig:spot_results} by noticing that the maximum value of the arches are situated closer to D1 for smaller SiPMs, which require a smaller spot size to reach these maximum values. In Figure \ref{fig:sipms_with_lens} an asymmetry in magnitude of the maxima on the arches and their distance to D1 is also observed. This is explained by the contribution of each diffraction order to the transmitted light. For every SiPM, the distance between the maximum value of the first arch (closer to the metalens) and D1 is greater than the distance between D1 and the maximum value of the second arch (further from the metalens). From Figures \ref{fig:lin_eff_results} and \ref{fig:foci_schematic}, it is observed that for every SiPM the maximum value of the first arch includes significant contributions from all diffraction orders, while the contributions of the 2$^{\text{nd}}$ and 3$^{\text{rd}}$ diffraction orders are reduced for the maximum value of the second arch, resulting in a reduced maximum value for the second arch at a location closer in distance to D1 relative to the maximum value of the first arch. Lastly, the effect of the multiple foci contributions on the signal shape is also noticeable in Figure \ref{fig:sipms_with_lens}, most notably for the $3\times3~\mathrm{mm}^2$ SiPM, where dips around 17 mm and 27 mm are observed due to the 3$^{\text{rd}}$ and 2$^{\text{nd}}$ order foci respectively, which are in agreement with Figure \ref{fig:spot_results}.

Figure \ref{fig:sipms_gain} shows the signal multiplication factor (SMF) as a function of distance from the metalens, which is defined as the amount of signal obtained when using a metalens (Figure~\ref{fig:sipms_with_lens}) divided the amount of signal obtained without using the metalens (Figure~\ref{fig:sipms_without_lens}) at every distance between each SiPM and the metalens. It is observed that the maximum SMF increases with decreasing SiPM size, reaching a value between 6 and 7 for the $1.3\times1.3~\mathrm{mm}^2$ SiPM. This provides insight into the effectiveness of projecting a larger metalens area onto a smaller SiPM area, as the ratio of the projected (metalens) area to the SiPM's active area increases with decreasing SiPM size. Moreover, the SMF curve for the $6\times6~\mathrm{mm}^2$ SiPM lies below 1.5 for all distances from the metalens, which demonstrates that metalenses are most effective when coupled to SiPMs with an active area that is at least a factor of 2 smaller than the area of the metalens. Lastly, the SMF for the $6\times6~\mathrm{mm}^2$ SiPM placed between 55~mm and 70~mm from the metalens falls below 1, indicating that the metalens actually hinders the light collection. 

\section{Conclusion}\label{sec:sec_conclusion}
Metalenses are a cost effective novel solution for improving the light collection efficiency of particle detectors with limited light detection coverage by light sensors such as SiPMs. The transmission properties of metalenses with a 10~mm diameter and a design wavelength of 632~nm are presented as a function of distance from the metalens. We demonstrate an increase in light collection by a factor between 6 and 7 when using a metalens to focus light onto a SiPM with an active area of $1.3\times1.3~\mathrm{mm}^2$, making metalenses most effective when coupled to light sensors of ever-smaller active areas. We provide a qualitative description of a light induced signal as a function of distance from the metalens when using metalenses with SiPMs, which is in agreement with the characterization of the transmitted light profile of the metalenses. While the results are obtained using a red LED, similar results are expected for metalenses with a design wavelength in the blue region of the electromagnetic spectrum (see Figure~1 in Ref.~\cite{Khorasaninejad1190}) due to similar light transmission efficiencies. Therefore, metalenses can be suitable for particle detectors that use wavelength shifters such as tetraphenyl butadiene (TPB). 

Further studies are needed to optimize the implementation of metalenses in a specific particle detector. For example, the array layout of metalenses and their properties such as design wavelength, size and shape, must be determined based on the characteristics of light sensors on a light collection plane. Modifications to the metalens design can also serve to converge the different foci to a single location in order to maximize the amount of light focused onto a single light sensor at a given distance from the metalens. Lastly, given the large interest in detection of vacuum-ultraviolet (VUV) light in noble element detectors, it would be worth investigating the design and fabrication of metalenses that can focus xenon or argon scintillation light.

\bibliographystyle{JHEP}
\bibliography{ref.bib}

\providecommand{\href}[2]{#2}\begingroup\raggedright\begin{thebibliography}{10}

\bibitem{Monrabal:2018xlr}
{\scshape NEXT} collaboration, \emph{{The NEXT-White (NEW) Detector}},
  \href{https://doi.org/10.1088/1748-0221/13/12/P12010}{\emph{JINST} {\bfseries
  13} (2018) P12010} [\href{https://arxiv.org/abs/1804.02409}{{\ttfamily
  1804.02409}}].

\bibitem{Alvarez:2012sma}
{\scshape NEXT} collaboration, \emph{{NEXT-100 Technical Design Report (TDR):
  Executive Summary}},
  \href{https://doi.org/10.1088/1748-0221/7/06/T06001}{\emph{JINST} {\bfseries
  7} (2012) T06001} [\href{https://arxiv.org/abs/1202.0721}{{\ttfamily
  1202.0721}}].

\bibitem{Albert:2017hjq}
{\scshape nEXO} collaboration, \emph{{Sensitivity and Discovery Potential of
  nEXO to Neutrinoless Double Beta Decay}},
  \href{https://doi.org/10.1103/PhysRevC.97.065503}{\emph{Phys.\ Rev.\ C}
  {\bfseries 97} (2018) 065503}
  [\href{https://arxiv.org/abs/1710.05075}{{\ttfamily 1710.05075}}].

\bibitem{Aalseth:2020nwt}
{\scshape DarkSide-20k} collaboration, \emph{{Design and Construction of a New
  Detector to Measure Ultra-Low Radioactive-Isotope Contamination of Argon}},
  \href{https://doi.org/10.1088/1748-0221/15/02/P02024}{\emph{JINST} {\bfseries
  15} (2020) P02024} [\href{https://arxiv.org/abs/2001.08106}{{\ttfamily
  2001.08106}}].

\bibitem{Aalbers:2016jon}
{\scshape DARWIN} collaboration, \emph{{DARWIN: towards the ultimate dark
  matter detector}},
  \href{https://doi.org/10.1088/1475-7516/2016/11/017}{\emph{JCAP} {\bfseries
  11} (2016) 017} [\href{https://arxiv.org/abs/1606.07001}{{\ttfamily
  1606.07001}}].

\bibitem{Khorasaninejad1190}
M.~Khorasaninejad, W.~T. Chen, R.~C. Devlin, J.~Oh, A.~Y. Zhu and F.~Capasso,
  \emph{Metalenses at visible wavelengths: Diffraction-limited focusing and
  subwavelength resolution imaging},
  \href{https://doi.org/10.1126/science.aaf6644}{\emph{Science} {\bfseries 352}
  (2016) 1190}.

\bibitem{Kildishev1232009}
A.~V. Kildishev, A.~Boltasseva and V.~M. Shalaev, \emph{Planar photonics with
  metasurfaces}, \href{https://doi.org/10.1126/science.1232009}{\emph{Science}
  {\bfseries 339} (2013) }.

\bibitem{Yu333}
N.~Yu, P.~Genevet, M.~A. Kats, F.~Aieta, J.~Tetienne, F.~Capasso et~al.,
  \emph{Light propagation with phase discontinuities: Generalized laws of
  reflection and refraction},
  \href{https://doi.org/10.1126/science.1210713}{\emph{Science} {\bfseries 334}
  (2011) 333}.

\bibitem{1514255}
W.~T. Chen, A.~Y. Zhu and F.~Capasso, \emph{Flat optics with
  dispersion-engineered metasurfaces},
  \href{https://doi.org/10.1038/s41578-020-0203-3}{\emph{Nature Reviews
  Materials} (2020) }.

\bibitem{Devlin10473}
R.~C. Devlin, M.~Khorasaninejad, W.~T. Chen, J.~Oh and F.~Capasso,
  \emph{Broadband high-efficiency dielectric metasurfaces for the visible
  spectrum}, \href{https://doi.org/10.1073/pnas.1611740113}{\emph{Proceedings
  of the National Academy of Sciences} {\bfseries 113} (2016) 10473}.

\bibitem{Park19}
J.-S. Park, S.~Zhang, A.~She, W.~T. Chen, P.~Lin, K.~M.~A. Yousef et~al.,
  \emph{All-glass, large metalens at visible wavelength using deep-ultraviolet
  projection lithography},
  \href{https://doi.org/10.1021/acs.nanolett.9b03333}{\emph{Nano Letters}
  {\bfseries 19} (2019) 8673}.

\bibitem{doi:10.1021/acs.nanolett.7b03135}
B.~H. Chen, P.~C. Wu, V.-C. Su, Y.-C. Lai, C.~H. Chu, I.~C. Lee et~al.,
  \emph{Gan metalens for pixel-level full-color routing at visible light},
  \href{https://doi.org/10.1021/acs.nanolett.7b03135}{\emph{Nano Letters}
  {\bfseries 17} (2017) 6345}.

\bibitem{doi:10.1021/nl302516v}
F.~Aieta, P.~Genevet, M.~A. Kats, N.~Yu, R.~Blanchard, Z.~Gaburro et~al.,
  \emph{Aberration-free ultrathin flat lenses and axicons at telecom
  wavelengths based on plasmonic metasurfaces},
  \href{https://doi.org/10.1021/nl302516v}{\emph{Nano Letters} {\bfseries 12}
  (2012) 4932}.

\bibitem{o2004diffractive}
D.~O'Shea, T.~Suleski, A.~Kathman and D.~Prather, \emph{Diffractive Optics:
  Design, Fabrication, and Test}, vol.~62 of \emph{SPIE, Bellingham}. 2004.

\bibitem{doi:10.1021/acsphotonics.9b00221}
M.~Decker, W.~T. Chen, T.~Nobis, A.~Y. Zhu, M.~Khorasaninejad, Z.~Bharwani
  et~al., \emph{Imaging performance of polarization-insensitive metalenses},
  \href{https://doi.org/10.1021/acsphotonics.9b00221}{\emph{ACS Photonics}
  {\bfseries 6} (2019) 1493}.

\end{thebibliography}\endgroup

\end{document}